\documentclass[prl,twocolumn,superscriptaddress,showpacs]{revtex4}
\input epsf

\begin{document}

\title{How One Shutter  Can Close $N$ Slits}

\author{Y. Aharonov}
\affiliation{ School of Physics and Astronomy\\ 
Raymond and Beverly Sackler Faculty of Exact Sciences \\
Tel-Aviv University, Tel-Aviv 69978, Israel}
\affiliation{ Physics Department, University of South Carolina\\
Columbia, South Carolina 29208, USA}
\author{L. Vaidman}
\affiliation{ School of Physics and Astronomy\\ 
Raymond and Beverly Sackler Faculty of Exact Sciences \\
Tel-Aviv University, Tel-Aviv 69978, Israel}

\date{\today}

\begin{abstract}
  It is shown that a quantum shutter, pre- and post-selected in
  particular quantum states, can close simultaneously arbitrary number
  of slits preventing the passage of a single photon in an arbitrary
  state. A set of $K$ pre- and post-selected shutters can close the
  slits preventing the passage of $K$ or less photons.  This result
  indicates that the surprising  properties of pre- and post-selected
  quantum systems are even more
  robust than previously expected.
\end{abstract}

\pacs{03.65.Ta}

\maketitle

  Probably the most paradoxical claim of quantum theory is that a
  particle can be in some sense in several places simultaneously.
  Without this feature one cannot explain the interference picture
  obtained in multiple-slit experiments performed with one particle at
  a time. A more robust claim of this type can be made about quantum
  pre- and post-selected particle. With utilization of  a particular pre- and
  post-selection, we can claim that the particle should have been
  found with certainty in each one out of several places given that it
  was looked for only in that place \cite{AV91}. Such claims
   became a subject of a
  significant controversy
  \cite{SS,DTSQT,Grif,Kent,Kast,reply,SHI,CO,CO-co}.  Here we discuss
  another aspect of such pre- and post-selected particle which makes
  the claim that such a particle is simultaneously in several places
  even more robust.

  Consider a photon arriving at a screen with $N$ holes (slits) at
  time $t$, Fig. 1.  We have a single particle (shutter) which, if
  placed in a slit, prevents the passage of the photon through this
  slit. Our task is to close all $N$ slits at time $t$ with this single
  shutter.

  We are allowed to
perform pre- and post-selection on the shutter: to prepare it at time
$t_1$ in state $|\Psi_1\rangle$ and select it in the state
$|\Psi_2 \rangle$ at $t_2$, $t_1 < t < t_2$. If the post-selection
measurement of $|\Psi_2\rangle$ fails, the experiment fails, but if it
succeeds, we should be able to claim that all $N$ slits were closed
for the photon at time $t$.

If the photon, bouncing of the shutter causes a measurable recoil,
then a post-selection can achieve this goal in a trivial way. We
just observe   the shutter
at time $t_2$. If we find a recoil, we may claim that the slits were
closed at time $t$. Indeed, we know that in this case the photon
bounced back from  the screen. However, in the present work we do not rely on
this effect. The setup is such that there is no measurable recoil.

The existence of a solution for this problem is surprising. The
probability for the photon to pass through the screen with the shutter in one
slit, or in an arbitrary superposition in all $N$ slits, is $1- 1/N$.
Nevertheless, the  pre- and post-selected shutter reflects the
photon with certainty.

To achieve this task we prepare (pre-select) the shutter at $t_1$ in
the state:
\begin{equation}
  \label{eq:psi1}
 |\Psi_1\rangle = {1\over{ \sqrt{2N-1}}}\left(\sum_{i=1}^{N} |i\rangle +
   \sqrt{N-1}~|N+1\rangle\right).
\end{equation}
We post-select the particle at $t_2$ in the state:
\begin{equation}
  \label{eq:psi2}
 |\Psi_2\rangle = {1\over{ \sqrt{2N-1}}}\left(\sum_{i=1}^{N} |i\rangle -
   \sqrt{N-1}~|N+1\rangle\right),
\end{equation}
where $|i\rangle$ is a state  of a shutter localized in  slit $i$, $i=1,
...,N$  and $|N+1\rangle$ is a state of the shutter
localized in  some specific different location.
 
In order to prove our claim, let us consider the time evolution of the
quantum state of the shutter and the photon during the whole
procedure. We assume that the free evolution of the shutter between
$t_1$ and $t_2$ can be neglected.
 Initially, the photon moves toward $N$ slits, so its state
is:

\begin{center} \leavevmode \epsfbox{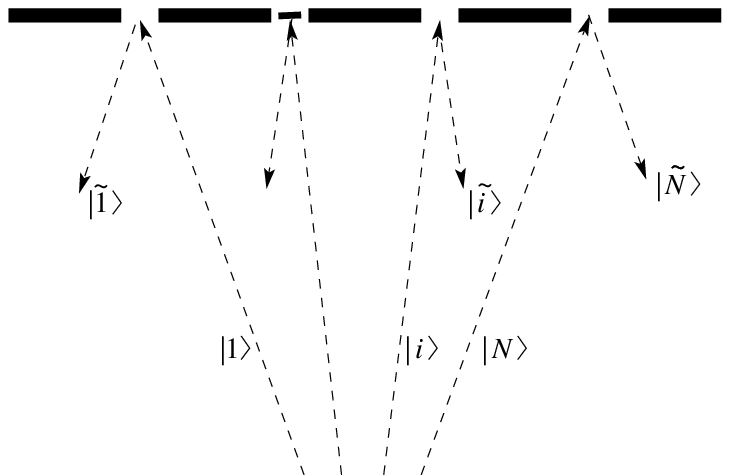} \end{center}
\noindent 
{\small {\bf Fig.~1.} A single photon arrives at $N$ slits, but a
  single shutter reflects the photon as if there were shutters in
  every slit. }

\break

\noindent
\begin{equation}
  \label{eq:psiPhin}
 |\Psi_{in}\rangle_{\rm ph}  = \sum_{i=1}^{N} \alpha_i |i\rangle_{\rm ph},
\end{equation}
where $|i\rangle_{\rm ph}$ is the state of a photon moving
toward  the slit $i$. Let us signify the state of a photon reflected
from slit $i$ as  $|\tilde i\rangle_{\rm ph}$. Then, after $t$,
the time of the interaction between the shutter and the photon, their joint
quantum state is:

\begin{widetext}
\begin{eqnarray}
  \label{eq:t1tt2}
\nonumber
 |\Psi\rangle_{\rm s,ph} =  {1\over{ \sqrt{2N-1}}}\sum_{i=1}^{N} |i\rangle \left( \alpha_i |\tilde
i\rangle_{\rm ph} +  \sum_{j\neq i}^{N}  \alpha_j | j\rangle_{\rm ph}\right) +
  \sqrt{{N-1}\over{2N-1}} ~|N+1\rangle \sum_{j=1}^N  \alpha_j  |j\rangle_{\rm ph} =\\
 {1\over{ \sqrt{2N-1}}} \sum_{i=1}^{N} \alpha_i |i\rangle |\tilde
 i\rangle_{\rm ph} +
 {1\over{ \sqrt{2N-1}}} \sum_{j=1}^{N} \alpha_j \left( \sum_{i\neq j}^{N}  |i\rangle +  \sqrt{N-1} ~|N+1\rangle \right)
 |j\rangle_{\rm ph} .
\end{eqnarray}
\end{widetext}

We can see that all states of the shutter appearing in the second term
in the last expression (i.e.,
all states correlated  with a photon which passed through the
screen) are orthogonal to the post-selected state
$|\Psi_2\rangle$. Therefore, after the post-selection, the photon
state will have only  reflected wave components. The screen  operates as a perfect mirror; the final
state of the photon is:
\begin{equation}
  \label{eq:psiPhout}
 |\Psi_{fin}\rangle_{\rm ph}  = \sum_{i=1}^{N} \alpha_i |\tilde i\rangle_{\rm ph}.
\end{equation}

We have shown that a single quantum shutter that has been pre- and
post-selected can close any number of slits.  It acts on the single
photon exactly in the same way as $N$ shutters.  Conceptually, using
this method one can build the whole screen out of a single pre- and
post-selected shutter (particle). This screen will act on a single
photon as a real screen made from many particles. In particular, a
photon passing through such a screen will follow a corresponding
diffraction pattern.

Not less surprising is a 
``dual'' problem which can be solved using our method. We have now $N$
shutters  which close   at least $N-1$ out of the $N$
slits. Nevertheless, we can  pre- and post-select the state of
these shutters in such a way that  a single photon will 
``see'' $N$ open slits. 

Consider the pre-selected  state of $N$ shutters
\begin{equation}
  \label{eq:psi1s}
 |\Phi_1\rangle = {1\over{ \sqrt{2N-1}}}\left(\sum_{i=1}^{N}|{\rm op}\rangle_i \prod_{j\neq i}^N|{\rm cl}\rangle_j  +
   \sqrt{N-1}\prod_{j=1}^N|{\rm cl}\rangle_j\right),
\end{equation}
where $|{\rm op}\rangle_i$  and $|{\rm cl}\rangle_i$   are the
states of a shutter corresponding to  an open
or  closed slit $i$  respectively. If now we test the number of
 closed slits, we will find with probability  ${{N-1}\over {2N-1}}$ that all
   slits are closed, and with probability  ${{N}\over {2N-1}}$ that all  but one slits
   are closed. However, we do not test the number of closed slits. We send
   at time $t$ the photon  in an arbitrary state (\ref{eq:psiPhin})
   toward the screen. Then, at time $t_2$,  we post-select the shutters  in the state:
\begin{equation}
  \label{eq:psi1s}
 |\Phi_2\rangle = {1\over{ \sqrt{2N-1}}}\left(\sum_1^{N}|{\rm op}\rangle_i \prod_{j\neq i}^N|{\rm cl}\rangle_j  -
   \sqrt{N-1}\prod_{j=1}^N|cl\rangle_j\right).
\end{equation}
A calculation, identical to the one performed above, 
shows that a single photon
passes the slits  without distortion, as if no shutters were present.

In our method a single (pre- and post-selected) shutter closes $N$
slits for a single photon. What will happen if at time $t$ several 
photons are trying to pass through the slits? If $K$ photons move toward the
screen in a particular correlated state
\begin{equation}
  \label{eq:prod}
 |\Psi\rangle_{K \rm ph} = \sum_{i=1}^{N} \alpha_i \prod_{k=1}^K
 |i\rangle_k , 
\end{equation}
then the shutter will reflect with certainty all the photons as it
reflected one. However, when the photons arrive in an arbitrary state, we
cannot be sure that even one photon will be reflected. Indeed, consider
an incoming two-photon state:
\begin{equation}
  \label{eq:prod2}
 |\Psi\rangle_{2\rm ph} = {1\over \sqrt 2} (|1\rangle_1 |2\rangle_2 +
 |2\rangle_1 |1\rangle_2).
\end{equation}
After the interaction between  the shutter and the photons at time $t$, the
state of the shutter correlated with the undisturbed state (\ref{eq:prod2})
is:
\begin{equation}
  \label{eq:prod2-}
  {1\over{ \sqrt{2N-3}}}\left(\sum_{i=3}^{N} |i\rangle +
   \sqrt{N-1}~|N+1\rangle\right).
\end{equation}
This state is not orthogonal to the post-selected state
(\ref{eq:psi2}). Therefore, a successful post-selection is possible
when both photons pass through the slits undisturbed, i.e., the two
photons might pass through the screen with our pre- and post-selected shutter.

In order to close $N$ slits for  a pair of photons we need {\it two}
pre- and post-selected shutters placed one after the other. The first
shutter should be pre-selected at time $t_1$ in the state:
\begin{equation}
  \label{eq:psi1'}
 |\Psi'_1\rangle  = {1\over{ \sqrt{2N-2}}}\left(\sum_{i=1}^{N} |i\rangle +
   \sqrt{N-2}~|N+1\rangle\right),
\end{equation}
 and post selected at time $t_2$ in the state:
\begin{equation}
  \label{eq:psi2'}
 |\Psi'_2\rangle = {1\over{ \sqrt{2N-2}}}\left(\sum_{i=1}^{N} |i\rangle -
   \sqrt{N-2}~|N+1\rangle\right).
\end{equation}

If the two photons pass through  two different slits without
disturbance, then the state  of the shutter will be  orthogonal to  $|\Psi'_2\rangle$.
 Therefore, given a successful
post-selection,  one photon should be reflected by the first shutter. 
 The second shutter is pre- and post-selected as in previous example,  in the states  (\ref{eq:psi1}) and 
 (\ref{eq:psi2}). This ensures reflection of the second photon.
 
 If the pair of photons pass through the same slit, then the state of
 the first shutter will not be orthogonal to $|\Psi'_2\rangle$. Therefore
 the photons in such a pair  might both pass through. But, in this case, the second
 shutter will reflect both photons with certainty, since it stops any number
 of photons arriving  together as in the correlated state (\ref{eq:prod}).

  In order to stop three photons we
 have to add another shutter in front of the  two described above. The additional shutter should reflect one photon any time three
 photons arrive at different slits. To this end, the shutter  should be
 pre- and post selected in the states $|\Psi''_1\rangle$ and
 $|\Psi''_2\rangle$:
\begin{equation}
  \label{eq:psi1''}
 |\Psi''_1\rangle  = {1\over{ \sqrt{2N-3}}}\left(\sum_{i=1}^{N} |i\rangle +
   \sqrt{N-3}~|N+1\rangle\right),
\end{equation}
\begin{equation}
  \label{eq:psi2'}
 |\Psi''_2\rangle = {1\over{ \sqrt{2N-3}}}\left(\sum_{i=1}^{N} |i\rangle -
   \sqrt{N-3}~|N+1\rangle\right).
\end{equation}
The generalization for larger number of photons is obvious. In this
way $K$ pre- and post-selected shutters close an arbitrarily large number
of slits $N$ for passing $K$ or less photons in an arbitrary state.

In this paper we have shown a surprising feature of pre- and
post-selected shutters. A single shutter can close an arbitrary number
of slits preventing the passage of a single photon in an arbitrary
state, while $K$ shutters can close the slits preventing  passage of any number of
photons $n \leq K$. On the other hand, $N$ shutters which close at
least $N-1$ slits can leave all slits open for a single photon.

For a pre- and post-selected state of a single shutter which closes
$N$ slits, it was known before \cite{AV91} that the outcomes of {\it
  weak measurements} performed in all slits correspond to one shutter
being in every slit. The present result shows that a measuring device,
namely the photon,
performing strong measurement while being in a superposition  in different
slits also indicates  the presence of the shutter in every
slit.

It is a pleasure to thank Shmuel Nussinov for helpful discussions.
This research was supported in part by grant 62/01 of the Israel
Science Foundation, by the Israel MOD Research and Technology Unit, by
the NSF grant PHY-9971005, and by the ONR grant N00014-00-0383.


\begin{thebibliography}{9}

\bibitem{AV91}
Y. Aharonov  and L. Vaidman, 
 J.  Phys.  A {\bf 24}, 2315 (1991).


\bibitem{SS}
W. D. Sharp  and N. Shanks,
 Phil.  Sci. {\bf 60}, 488 (1993).

\bibitem{DTSQT} 
L. Vaidman,
 Stud. Hist. Phil. Mod. Phys. {\bf 30}, 373 (1999). 



 \bibitem{Grif}
R. B. Griffiths, Phys. Rev. A {\bf  54}, 2759 (1996);
     {\bf 57}, 1604 (1998).

 \bibitem{Kent}
A. Kent, Phys. Rev. Lett. {\bf 78}, 2874 (1997). 





\bibitem{Kast}
R. E. Kastner,
Found. Phys. {\bf 29}, 851 (1999).



\bibitem{reply}
L. Vaidman,
Found. Phys. {\bf 29}, 865 (1999).


\bibitem{SHI}
A. Shimony,
 Erken. {\bf 45}, 337 (1997).


\bibitem{CO}
O. Cohen,
  Phys.  Rev. A  {\bf 51},  4373 (1995). 
 
\bibitem{CO-co}
L. Vaidman,
 Phys.  Rev. A {\bf  57}, 2251 (1998).

       
\end{thebibliography}
\end{document}